\documentclass[%
 aip,
 jmp,%
 amsmath,amssymb,
 reprint,%
]{revtex4-1}

\usepackage{graphicx}
\usepackage{dcolumn}
\usepackage{bm}
\usepackage{color}
\usepackage{float}

\begin{document}


\title{Brightness of nonequilibrium photoemission in metallic photocathodes near threshold
}

\author{Jai Kwan Bae}
\email{jb2483@cornell.edu}
\author{Ivan Bazarov}

\affiliation{Cornell Laboratory for Accelerator-Based Sciences and Education, Cornell University, Ithaca, NY 14853, USA}

\author{Pietro Musumeci}
\affiliation{Department of Physics and Astronomy, UCLA, Los Angeles, CA 90095, USA}

\author{Siddharth Karkare}
\affiliation{Department of Physics, Arizona State University, Tempe, AZ 85287, USA}

\author{Howard Padmore}
\affiliation{Advanced Light Source Division, LBNL, Berkeley, CA 94720, USA}

\author{Jared Maxson}
\email{jmm586@cornell.edu}
\affiliation{Cornell Laboratory for Accelerator-Based Sciences and Education, Cornell University, Ithaca, NY 14853, USA}

\date{\today}

\begin{abstract}
The operation of photoemission electron sources with wavelengths near the photoemission threshold has been shown to dramatically decrease the minimum achievable photocathode emittance, but at the cost of significantly reduced quantum efficiency (QE). In this work, we show that for femtosecond laser and electron pulses, the increase in required laser intensities due to the low QE drives the photocathode electronic distribution far from static equilibrium. We adapt an existing dynamic model of the electron occupation under high intensity laser illumination to predict the time-dependent effects of the nonequilibrium electron distribution on the QE, mean transverse energy (MTE), and emission brightness of metal phtocathodes. We find that multiphoton photoemission dramatically alters the MTE as compared to thermal equilibrium models, causing the MTE to no longer be a monotonic function of photon excess energy. 

\end{abstract}

\maketitle

\section{Introduction}
Ultrafast  electron pulses generated from photocathode sources play a critical role in probing time-resolved ultrafast dynamics, in applications ranging from free electron lasers\cite{fel} to ultrafast electron diffraction\cite{ued} and microscopy.\cite{uem} For each of these cases, the critical figure of merit is the 5D beam brightness. In a linear accelerator, the beam brightness is never larger than it is at the photoemission source, and is thus given by:\cite{Rhee1992,ttm,pancake}
\begin{equation}\label{eq_brightness}
\mathcal{B}_{5D, max} \equiv \frac{1}{16 \pi^2} \frac{I}{\epsilon_x \epsilon_y} = \frac{1}{16 \pi^2} \frac{I m_e c^2}{\sigma_x \sigma_y \text{MTE}} 
\end{equation}
where $I$ is the peak current from the cathode, $\epsilon_{x,y}$ is the normalized emittance of the beam, $\sigma_{x,y}$ is the rms laser spot size, $m_e c^2$ is the electron rest energy, and MTE is the mean transverse energy of the photoelectrons at the photocathode.\cite{Musumeci2018} The laser spot size is primarily determined by the space charge beam dynamics after photoemission. The MTE is determined by material properties that affect the photoemission process inside the photocathode, the cathode surface, and the photoemission drive laser properties.



Dowell and Schmerge demonstrated that Spicer's three-step photoemission model\cite{spicer} can be utilized to derive expressions for the photoemission quantum efficiency (QE) and MTE of metal photocathodes.\cite{3step} While this was originally derived under the assumption of a free electron Fermi gas and a flat density of states at zero temperature,  recent studies have extended this approach to account for a realistic density of states and non-zero (but constant) electronic temperature ($T_e$). \cite{Dimitrov2017,Feng2015} These models predict that as the photon energy ($h\nu$) approaches the work function ($\phi$), MTE converges to the thermal energy of electrons (MTE $\approx k_B T_e$). For a photon energy well above the threshold ($h\nu - \phi \gg 0$), the MTE is linear in the photon excess energy, MTE $ \approx (h\nu-\phi)/3$. Both behaviors have been demonstrated experimentally.\cite{Feng2015}  These studies predict that MTE is minimized by operating with very low photon excess energy, but do not account for dynamic or intensity dependent effects, such as multiphoton excitation or laser-induced ultrafast electron heating.


Metallic photocathodes, and in particularly copper photocathodes, are popular choices for ultrafast electron sources due to their prompt response time ($<50$ fs) and moderate vacuum requirements.\cite{ttm,plasmon} However, compared to semiconducting emitters, they suffer from several orders of magnitude less QE, due to frequent electron-electron scattering  during the transport of the excited population from the bulk to the surface. Thus, metallic photocathodes can demand a very high laser intensity   ($\text{10s of GW/cm}^2$) for femtosecond emission cases, for example in the blowout regime\cite{blowout} of high charge photoinjectors.

It has long been known that under high fluence, short pulse laser irradiation, the electronic temperature of the illuminated material can reach several thousand Kelvin for several picoseconds, while the lattice remains relatively cold.\cite{Anisimov1974,Fann1993,Elsayed-Ali1987,Bartoli1997} It was recently pointed out that this effect can drive up the minimum achievable photoemission MTE, and was calculated by using the traditional two temperature model coupled with the extended Dowell-Schmerge photoemission relations.\cite{ttm} Two temperature models naturally assume the underlying electronic distribution remains approximately thermalized, which may not be the case for femtosecond duration illumination. Furthermore, there have been recent experimental efforts which highlight the importance of multiphoton and heating effects on intense femtosecond photoemission.\cite{Pasmans2016,An2018}

Rethfeld \textit{et al.} used the Boltzmann equation to study femtosecond nonequilibrium thermodynamics of metals irradiated by high laser fluence.\cite{rethfeld,boltzmann,skin}
It was shown that this approach also allows simulating dynamic effects of photoemission, such as the image-charge driven reduction of the work function via a time dependent space charge field.\cite{rethfeld_schottky}
In this paper, we extend the Rethfeld approach to calculate key brightness parameters (QE and MTE) of a photocathode under high fluence ($\sim 1 \text{ mJ/cm}^2$) laser irradiation for femtosecond electron pulse applications. The model has the capability to account for femtosecond nonequilibrium thermodynamics, multiphoton absorption, and a realistic density of states.



\section{Boltzmann equation}

Before photons arrive at the cathode, the initial electronic occupation function, $f(\vec{k})$, is in thermal equilibrium, and thus follows the Fermi-Dirac distribution. As photons begin to interact with electrons, the electronic occupation changes as a function of time, and we can calculate this change for each time step according to the Boltzmann equation:
\begin{equation}
\frac{d f(\vec{k},t)}{d t} = \frac{\partial f(\vec{k},t)}{\partial t}\biggr\rvert_{el-el} +  \frac{\partial f(\vec{k},t)}{\partial t}\biggr\rvert_{\substack{photon \\ absorb}}
\label{bolt}
\end{equation}
where the first and second terms represent the occupation function change due to electron-electron collisions and photon absorption respectively. Here and throughout we have neglected the effects of electron-phonon interactions due to the very short timescale ($\leq 100$ fs) of the laser pulses considered. Furthermore, thermal conductivity and all spatial variation of the laser intensity is ignored, which is valid only for the case of homogeneous thin film photocathodes.\cite{boltzmann}

Whereas the photon absorption term in Eq.~\ref{bolt} drives the distribution away from thermal equilibrium, the electron-electron scattering term regulates the thermalization timescale. As in Ref.~\onlinecite{boltzmann}, the electron-electron scattering term is calculated based on the Coulomb interaction of two electrons that obey energy conservation ($\Delta E = E_2 - E_1 = E_3 - E$) and momentum conservation ($\Delta \vec{k} = \vec{k_2} - \vec{k_1} = \vec{k_3} - \vec{k}$).\cite{rethfeld,boltzmann} Here, $\vec{k}, \vec{k_2}$ are initial wave vectors, and $\vec{k_1}, \vec{k_3}$ are final wave vectors of the two interacting electrons. Assuming an isotropic system, the Fourier transform of the screened Coulomb potential $V_{ee}(r)$ yields:\cite{Ashcroft1978,rethfeld}
\begin{equation}\label{eq_m}
|M_{ee}(\Delta k,\kappa)|^2 = \Big(\frac{e^2}{\epsilon_0 \Omega}\frac{1}{|\Delta k|^2 + \kappa^2}\Big)^2
\end{equation}
where $\Omega$ is the unit cell volume, and
\begin{equation}
\kappa^2 (t) = \frac{e^2}{\epsilon_0}\int_0^\infty dE D(E) \frac{df(E,t)}{dE}
\end{equation}
is the screening parameter with $D(E)$ as the density of state per volume for copper.\cite{Lin2008} The density of states is assumed to be flat for energy 6 eV above the Fermi level. Then, due to Fermi's golden rule, the change in the occupation due to electron-electron collision processes is given by:\cite{boltzmann,rethfeld}
\begin{equation}\label{eq_ee}
\frac{\partial f(\vec{k})}{\partial t}\biggr\rvert_{el-el} = \frac{2 \pi}{\hbar}\sum_{\vec{k_1}}\sum_{\vec{k_3}} |M_{ee}|^2 \mathcal{F} \delta(E_i-E_f).
\end{equation}
Here, $\mathcal{F}$ is the Pauli exclusion principle term expressed with occupation functions, $f_i \equiv f(k_i)$:\cite{rethfeld,boltzmann,skin}
\begin{equation}
\mathcal{F} = f_1 f_3 (1-f)(1-f_2) - f f_2(1-f_1)(1-f_3).
\end{equation}
In this work, we further assume the system to be described by an isotropic single band model, which in analogy with the free electron case yields:
\begin{equation}
|k(E)|^3 = 3\pi^2 \int_0^E d\epsilon D(\epsilon).
\end{equation} 
Consequently, Eq.~\ref{eq_ee} can be further simplified as in Ref. \onlinecite{boltzmann}.

The effective one band assumption requires a mediating body to conserve momentum upon photon absorption.  Here the inverse bremsstrahlung model is used: quasifree electrons absorb photons mediated by Coulomb potential from an ion. Here, the electron mass is negligible compared to the ion mass. Therefore, the momentum conservation holds for an arbitrary final momentum ($\vec{k} + \Delta \vec{k} =\vec{k_f}$) while there is no energy transfer to the ion. The Coulomb interaction between an electron and ion has mathematically identical expression to Eq.~\ref{eq_m}. Additionally, the interaction between the electron and photon introduces a Bessel function term, $J_\ell^2(e\vec{E_L}\cdot \Delta \vec{k}/m_e \omega_L^2)$, multiplied to the Coulomb potential term where $\ell$ is the number of photons absorbed, $\omega_L$ is the angular frequency of the laser light, and $\vec{E_L}$ is the electric field from the photons. Hence, the Fermi's golden rule is given by:
\begin{equation}\label{eq_absorb}
\frac{\partial f(\vec{k})}{\partial t}\biggr\rvert_{\substack{photon \\ absorb}} = \frac{2\pi}{\hbar}\sum_{\Delta\vec{k}}\sum_\ell |M_{ee}|^2 J_\ell^2 \mathcal{F} \delta(E_i-E_f)
\end{equation}
where $\mathcal{F}$ is again the Pauli exclusion principle term:\cite{rethfeld,boltzmann}
\begin{equation}
\mathcal{F} = f(k_f)(1-f(k))-f(k)(1-f(k_f)).
\end{equation}
Analogous to Eq.~\ref{eq_ee}, the simplified expression of Eq.~\ref{eq_absorb} from Ref.~\onlinecite{boltzmann} is used throughout in our work. It is important to note that this model allows for both the simultaneous (via the power series in $\ell$) and subsequent (time delayed) absorption/emission of multiple photons. However, the simultaneous, or coherent, multiphoton absorptions are suppressed in this model due to single band assumption.

\begin{figure}
   \includegraphics*[width=250pt]{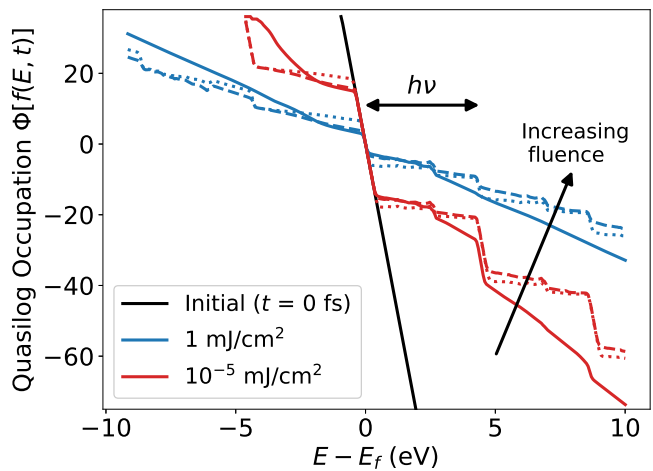}
   \caption{Quasilogarithmic transformation of electron energy distribution $\Phi[f(E,t)]$ of copper over time (Eq.~\ref{eq_quailog}).  Under 50 fs laser irradiation, the work function photon energy of 4.31 eV was used with two different laser intensities that results in 1 mJ/cm$^2$ (blue) and $10^{-5} \text{ mJ/cm}^2$ (red) absorbed fluences. The transient distributions are plotted at $t=$ 10 fs (dotted lines), 50 fs (dashed lines), and 100 fs (solid lines).}
   \label{fig_occupation}
\end{figure}

To investigate the dynamics of electron energy distribution of copper during and after 50 fs laser irradiation, we perform a quasilogarithmic transformation of the occupation function $f(E,t)$ as proposed in Ref.~\onlinecite{skin}:
\begin{equation} \label{eq_quailog}
\Phi[f(E,t)] \equiv -\ln[1/f(E,t)-1].
\end{equation}
Under this transformation, a Fermi-Dirac distribution has a linear relation with energy where its slope is inversely proportional to the temperature. Note that nonlinearity in energy indicates nonequilibrium behavior. In Fig.~\ref{fig_occupation}, a quasilogarithmic transformation of the occupation functions of copper under 50 fs laser irradiation is plotted for two different absorbed fluences at four different time positions (0 fs, 10 fs, 50 fs, and 100 fs from the beginning of the laser irradiation). The initial distribution demonstrates a perfect linear relation in thermal equilibrium of 300 K whereas the occupation functions during the laser irradiation ($0<t \leq 50$ fs) show strong nonlinearity due to nonequilibrium dynamics. At 100 fs, the high fluence irradiated electrons approach another thermal equilibrium of $\sim$3500 K while the low fluence irradiated electrons are still far from thermalization. It was reported that a smaller fluence irradiation tends to have a longer thermalization timescale up to several hundred femtoseconds.\cite{boltzmann}

\section{Calculation of QE and MTE in nonequilibrium conditions} 

The QE and MTE can be simulated by determining the number and energy distribution of electrons promoted above the vacuum level based on numerically calculated time dependent occupation function $f(E,t)$ from the previous section. The energy distribution of electrons above the vacuum level parameters is dynamic, meaning that the efficiency, MTE, and photoemitted pulse shape are nontrivial functions of time.

QE is defined as the number of escaped electrons divided by the number of photons in each laser pulse. To calculate the number of escaped electrons, we express the partial current density of electrons per energy $E$ in the direction $\theta$ away from the normal direction as:
\begin{equation}
j(E,t,\theta) \equiv e f(E,t) D(E) v(E) \cos\theta 
\end{equation}
where $e$ is the electron charge, and $v(E)$ is the speed of electrons ($\hbar k(E) / m_e$).
Then,  the total number of escaped electrons per unit area can be derived by integrating  this flux over time ($t$), energy above the vacuum level ($E > E_f + \phi$), and allowed directions of the escape velocity ($\cos \theta > \cos\theta_E \equiv \sqrt{(E_f + \phi)/E}$):\cite{3step}
\begin{widetext}
\begin{equation}\label{eq_Ne}
N_e = \frac{\displaystyle \int_0^t dt' \int_{E_f+\phi}^\infty dE \int^1_{\cos\theta_E} d(\cos\theta) \frac{j(E,t',\theta)}{e}}{\displaystyle \int^1_{-1}d(\cos\theta)}.
\end{equation}
\end{widetext}

Although it may be intuitive to calculate the number of photons in each pulse based on the input parameters from Eq.~\ref{eq_absorb}, it was shown that inverse bremsstrahlung process, which accounts only intraband absorption, does not produce the correct optical skin depth of the material.\cite{skin} The discrepancy can be up to a factor of 2 for the range of photon energy used in our work. Therefore, instead of using the input parameters, the number of absorbed photons is calculated \emph{post-facto} based on the change of internal energy for unit volume, $\Delta E$:
\begin{widetext}
\begin{equation}\label{eq_Nph}
N_{ph} = \frac{d_s \Delta E}{ \hbar \omega_L} = \frac{d_{s}}{\hbar \omega_L} \int_0^\infty dE \: D(E) [f(E,t)-f(E,0)] E.
\end{equation}
\end{widetext}
Here, the optical skin depth, $d_s$ = 13 nm,\cite{CuRefl} is multiplied to calculate the number of absorbed photons for unit area. Finally, the QE is expressed as:
\begin{equation}\label{eq_qe}
\text{QE} = N_e (1-R) / N_{ph}
\end{equation}
where $R$ is the reflectivity of copper.\cite{CuRefl}

MTE is the averaged transverse energy ($(p \sin\theta)^2/2m_e$) of electrons above the vacuum level. As in the Dowell-Schmerge model,  we assume no effects of the electron effective mass on the MTE, and thus  set $p^2/2m_e = E$. Similar to Eq.~\ref{eq_Ne}, we get MTE by integrating over time, energy, and allowed angle:
\begin{widetext}
\begin{equation}\label{eq_mte}
\textmd{MTE} = \frac{\displaystyle \int_0^t dt' \int_{E_f+\phi}^{\infty} dE \int_{\cos\theta_E}^1 d(\cos\theta) j(E,t',\theta) E \sin^2\theta}{\displaystyle \int_0^t dt' \int_{E_f+\phi}^{\infty} dE \int_{\cos\theta_E}^1 d(\cos\theta) j(E,t',\theta)}.
\end{equation}
\end{widetext}
Due to discretization of the energy distribution, the uncertainty of calculated MTE is $\sim$ 5 meV.

\section{Results: QE and MTE}

\begin{figure}[h!]
   \includegraphics*[width=250pt]{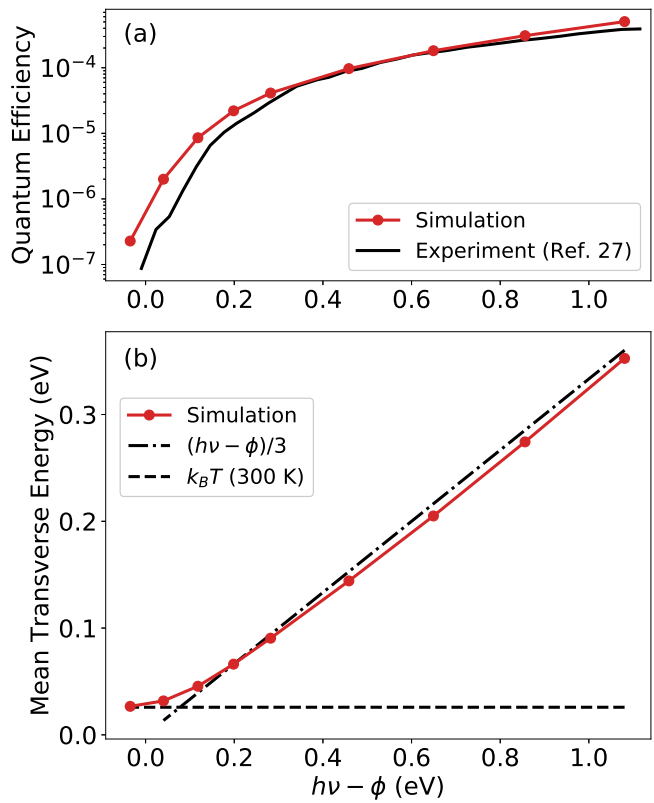}
   \caption{(a) QE calculated based on Eq.~\ref{eq_qe} in single photon absorption dominant regime ($10^{-5} \text{ mJ/cm}^2$ absorbed fluence), compared to experimentally measured values.\cite{Dowell2006} (b)  MTE computed by Eq.~\ref{eq_mte} as a function of excess photon energy ($h\nu - \phi$) for $10^{-5} \text{ mJ/cm}^2$ absorbed fluence. The dashed line represents the thermal energy. The dash-dotted line is a linear relation predicted and confirmed in earlier works.\cite{3step,Feng2015,Dimitrov2017}}
   \label{fig_low}
\end{figure}

We first set the absorbed laser fluence down to single-photon absorption dominant regime ($10^{-5} \text{ mJ/cm}^2$) to compare with experimentally measured QE \cite{Dowell2006} and earlier predictions on MTE.\cite{3step,Feng2015,Dimitrov2017} For a pulse length of 50 fs, the QE and MTE for a copper photocathode with a realistic density of states are plotted in Fig.~\ref{fig_low} based on Eq.~\ref{eq_qe} and Eq.~\ref{eq_mte}.
Throughout the calculations, 4.31 eV was used for the work function ($\phi$). The quantum efficiency demonstrates very good agreement with experimentally measured values from a copper photocathode cleaned by a hydrogen ion beam in situ.\cite{Dowell2006} Furthermore, the asymptotic properties of MTE is consistent with earlier predictions.\cite{3step,Feng2015,Dimitrov2017}

\begin{figure}
   \includegraphics*[width=250pt]{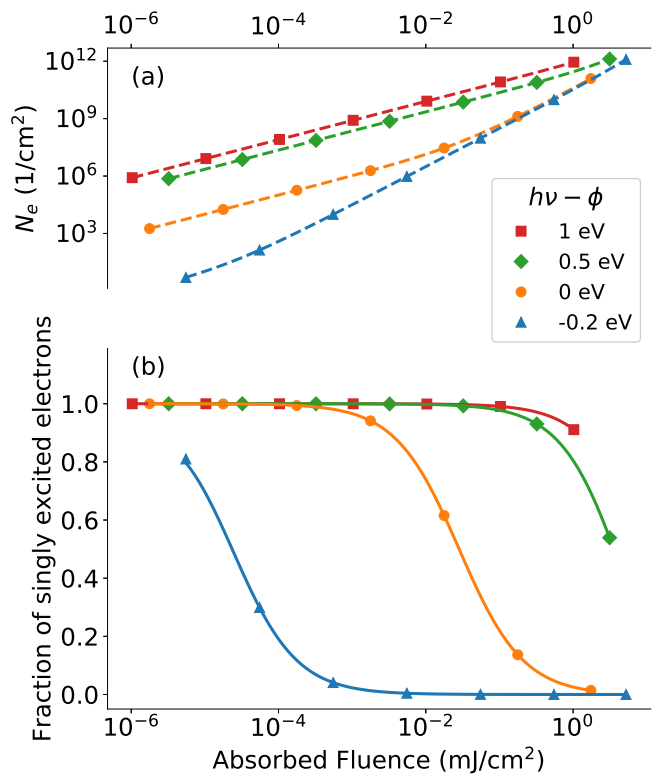}
   \caption{(a) The total number of extracted electrons in each pulse for unit area fitted to a fourth degree polynomial function of absorbed fluence. (b) Fraction of singly excited electrons calculated based on generalized Fowler-DuBridge theory using Eq.~\ref{eq_single}.}
   \label{fig_single}
\end{figure}

A generalized Fowler-DuBridge analysis was performed to investiagte the contribution of multiphoton effects in the QE and MTE. The generalized Fowler-DuBridge theory for multiphoton photoemission suggests that the total photocurrent density $J$ can be expressed as a sum of partial current densities due to $n$-photon absorption $J_n$:\cite{Bechtel1977,Ferrini2009,multiphoton_musumeci}
\begin{equation}
J = \sum_{n = 0}^{\infty} J_n = \sum_{n = 0}^{\infty} \sigma_n I^n
\end{equation}
where $J_n$ is proportional to the $n$-th power of the intensity of incident laser $I$, and $\sigma_{n}$ is the series coefficient. Hence, the fraction of singly excited electrons among all emitted electrons can be calculated as:\cite{An2018}
\begin{equation}\label{eq_single}
\frac{J_1}{J} = \frac{\sigma_1 I}{\sum_{n=0}^\infty \sigma_n I^n}.
\end{equation}
In Fig.~\ref{fig_single}, the total number of emitted electrons $N_e$ (Eq.~\ref{eq_Ne}) is fitted to a fourth degree polynomial function of absorbed fluence (Eq.~\ref{eq_Nph}), to plot the fraction of singly excited electrons above the vacuum level.
For $10^{-5} \text{ mJ/cm}^2$ absorbed fluence used in Fig.~\ref{fig_low}, single photon absorbed electrons dominate for all photon energies greater than the work function (See Fig.~\ref{fig_single} (b)). Thus, the calculated photoemission parameters are essentially determined by singly excited electrons and agree well with conventional, static models that only consider single photon absorption. However, as the absorbed fluence increases, the fraction of electrons that absorbs a single photon drops rapidly near the threshold ($h\nu - \phi \approx 0$), and multiphoton absorbed electrons are no longer negligible in calculation of QE and MTE.

\begin{figure}
   \includegraphics*[width=250pt]{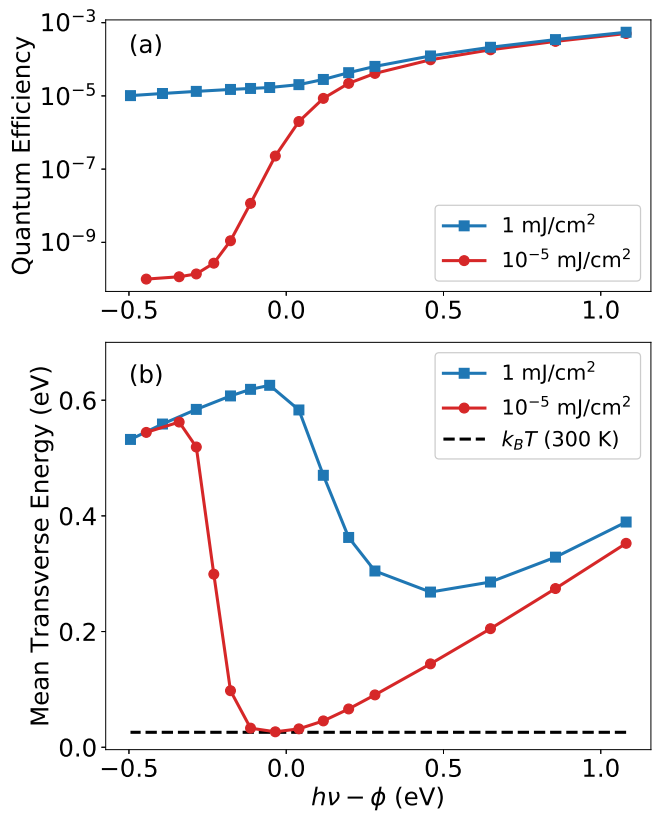}
   \caption{(a) QE computed by Eq.~\ref{eq_qe} for a high fluence of $1 \text{ mJ/cm}^2$, compared to the low fluence ($10^{-5} \text{ mJ/cm}^2$) simulation over extended photon energy range. (b) MTE calculated based on Eq.~\ref{eq_mte} for $1 \text{ mJ/cm}^2$ and $10^{-5} \text{ mJ/cm}^2$ absorbed fluences. Two-photon absorbed electrons cause the non-monotonic behavior of MTE.}
   \label{fig_high}
\end{figure}

To demonstrate the effect of using high fluence laser, QE and MTE are recalculated in Fig.~\ref{fig_high} with  $1 \text{ mJ/cm}^2$ absorbed fluence. For photon energies well above the threshold, it is noteworthy that both QE and MTE values are near those in the low fluence simulation. This is consistent with Fig.~\ref{fig_single}, as singly excited electrons account for the majority of electrons that have escaped even at high fluence with high photon excess energy.
On the other hand, as the photon energy approaches to the work function ($h\nu - \phi \approx 0$), the fraction of singly-excited emitted electrons rapidly decreases (See Fig.~\ref{fig_single}(b)). This tendency is also enhanced by the fact that the allowed direction of the escape velocity ($ \cos\theta > \sqrt{(E_f + \phi)/E}$) is substantially suppressed near the threshold energy ($E_f + \phi$) of electrons. Hence, two-photon absorption becomes non-negligible well above the work function photon energy and yields a non-monotonic behavior of MTE as a function of the photon excess energy ($h\nu - \phi$) in Fig.~\ref{fig_high}. If the photon energy falls below the work function, single photon absorption can no longer contribute electrons above the vacuum level, so the photoemission parameters are essentially dominated by two-photon absorption alone. Thus, once photoemission is entirely dominated by two photon effects, a further reduction of the photon excess energy once again yields a reduction of the MTE, but the thermal value is never achieved.

Recently, increase of intrinsic emittance induced by multiphoton photoemission from copper photocathodes under femtosecond laser illumination was experimentally verified,\cite{An2018} and showed qualitative agreement with our results. In that work, the growth of MTE was observed at the absorbed fluence of $7 \times 10^{-6} \text{ mJ/cm}^2$ with the photon energy 0.2 eV below the effective work function. According to our simulation in Fig.~\ref{fig_single}, the fraction of singly excited electrons is $\sim 76 \%$ for a similar laser profile.
When the photon energy range was extended below the threshold for the low fluence simulation in Fig.~\ref{fig_high}, we also observe the MTE growth.
Note the non-monotonic behavior of MTE can be captured theoretically only when multiphoton absorption is considered.

\section{Brightness}

\begin{figure}
   \includegraphics*[width=250pt]{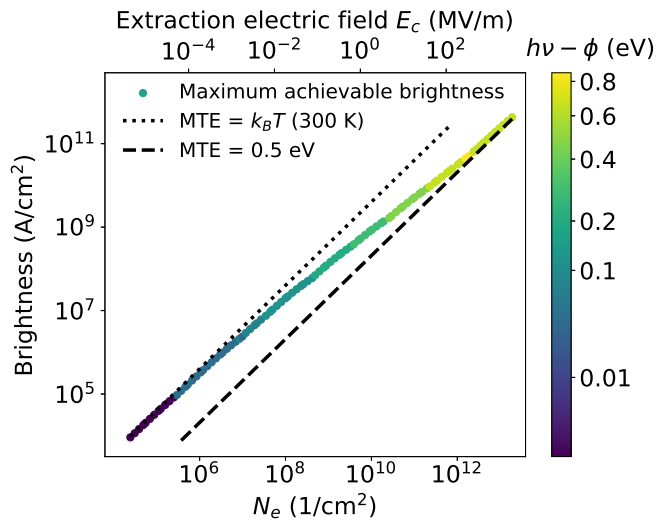}
   \caption{Maximum achievable brightness calculated by Eq.~\ref{eq_brightness} as a function of the number of extracted electrons from Eq.~\ref{eq_Ne} and the required extraction field under the pancake aspect ratio approximation.\cite{pancake} The color scale represent the input photon energy to obtain the corresponding maximum achievable brightness. The dotted line and dashed line illustrate brightness calculated using fixed MTE. One thousand sets of photon energy and laser fluence were simulated to acquire maximum achievable brightness.}
   \label{fig_brightness}
\end{figure}

One of notable aspects of the high fluence laser irradiation simulation in Fig.~\ref{fig_high} is that the laser photon energy can be tuned to minimize MTE, thereby maximizing brightness for a fixed laser fluence. Therefore, by iterating over numerous input laser fluences and photon energies, one can plot the maximum achievable brightness for a given number of extracted electrons per pulse and the corresponding required extraction field $E_c$ as in Fig.~\ref{fig_brightness}.
In the pancake aspect ratio regime, the charge density is directly proportional to the required extraction field, $e N_e = \epsilon_0 E_c$.\cite{pancake}
Here, the brightness and the number of electrons for a 50 fs flat top pulse are calculated by Eq.~\ref{eq_brightness} and Eq.~\ref{eq_Ne}, respectively. The color scale shows the optimal photon energy above the threshold used to obtain the maximum brightness for corresponding charge extraction per pulse. Two features are of note. First, the optimal photon energy can be well above the threshold depending on desired amount of charge extraction for specific applications. Second, the maximum brightness as a function of the extracted areal charge density is no longer linear, but rather at 100 MV/m equivalent space charge fields the maximum brightness is reduced below the room temperature limit by a factor of $\sim 12$.

\section{Conclusion}

We have presented photoemission simulations using a  Boltzmann equation method that has capability to account nonequilibrium thermodynamics and multiphoton photoemission. Using a low absorbed fluence ($10^{-5} \text{ mJ/cm}^2$), the calculated photoemission parameters reproduced the experimentally measured QE and demonstrated good agreement of MTE with earlier, static equilibrium predictions due to the abundance of singly excited electrons. In contrast, for a high absorbed fluence ($1 \text{ mJ/cm}^2$), multiphoton absorbed electrons are no longer negligible especially near the threshold photon energy, thereby causing the growth of MTE. Since the minimum MTE is no longer the thermal energy and not achieved by the work function photon energy, a series of laser fluences were simulated to plot the maximum achievable brightness for a given number of extracted electrons in a 50 fs pulse.  Our results illustrate the importance of multiphoton effects on beam brightness of photocathodes for femtosecond applications. A dynamic photoemission calculation like the one presented here could be self-consistently coupled to a space charge dynamics tracking code for precision photoinjector modeling. 




\section{Acknowledgments}
This work was supported by the U.S. National Science Foundation under award PHY-1549132, the Center for Bright Beams.

\bibliographystyle{unsrt}
\bibliography{boltzmann}

\end{document}